\documentclass[aps,prl,twocolumn,amsmath,amssymb,superscriptaddress,10pt]{revtex4-2}
\usepackage{fix-cm}
\usepackage{array}
\usepackage{bm}
\usepackage{color}
\usepackage{float}
\usepackage{graphicx}
\usepackage[caption=false]{subfig}

\usepackage{newtxtext}
\usepackage{newtxmath}

\usepackage{enumitem}
\usepackage{gensymb}
\usepackage{natbib}
\usepackage{soul}
\usepackage{lipsum}
\usepackage{hyperref}
\hypersetup{colorlinks=true,allcolors=blue}

\begin{document}

\title{Pair-Density Wave from Doping an Altermagnetic Mott Insulator}

\author{Shuai A. Chen}
\thanks{These authors contributed equally to this work}
\email{chsh@pks.mpg.de}
\affiliation{Max Planck Institute for the Physics of Complex Systems, N\"{o}thnitzer Stra{\ss}e 38, Dresden 01187, Germany}

\author{Qianqian Chen}
\thanks{These authors contributed equally to this work}
\affiliation{Kavli Institute for Theoretical Sciences and School of Quantum, University of Chinese Academy of Sciences, Beijing 100190, China}

\author{Guangyu Yu}
\affiliation{Kavli Institute for Theoretical Sciences and School of Quantum, University of Chinese Academy of Sciences, Beijing 100190, China}

\author{Zheng Zhu}
\email{zhuzheng@ucas.ac.cn}
\affiliation{Kavli Institute for Theoretical Sciences and School of Quantum, University of Chinese Academy of Sciences, Beijing 100190, China}

\date{\today}
\begin{abstract}
  Pair-density-wave (PDW) superconductivity is a state in which the superconducting
  order parameter modulates at a finite wavevector. Using large-scale density
  matrix renormalization group, we study the doped altermagnetic Mott insulator in the checkerboard $t$-$J$
  model, where altermagnetic exchange anisotropy is encoded microscopically through
  anisotropic ferromagnetic next-nearest-neighbor exchange.
  By mapping the ground-state phase diagram as a function of doping and altermagnetic anisotropy, mainly on six-leg cylinders, we identify a
  transition from a uniform $d$-wave superconducting regime with charge
  modulation to a PDW regime coexisting with stripe order. In the PDW regime, we
  report an unconventional wave-vector locking $\mathbf Q_{\mathrm{PDW}}\approx
    2\mathbf Q_{\mathrm{Stripe}}$ along the cylinder direction,
  in contrast to the conventional relation.
  Pair correlations reveal a two-scale structure, consisting of short-distance
  local $d$-wave pairing and long-distance finite-momentum PDW correlations. A
  symmetry-based Ginzburg--Landau analysis is presented for the observed locking.
  Our results identify altermagnetism as a strong-coupling, microscopically
  grounded route to finite-momentum superconductivity in doped Mott insulators. 
\end{abstract}

\maketitle

\emph{\color{blue}Introduction.}---The pair-density wave (PDW) is an unconventional superconducting state
in which the superconducting order parameter oscillates in real space at a finite wavevector $\mathbf{Q}_\mathrm{PDW}$~\cite{PhysRevX.4.031017,hamidian2016detection}.
In doped Mott insulators, holes often self-organize into charge stripes that separate remnant antiferromagnetic domains~\cite{RevModPhys.87.457,Agterberg2020}.
In the conventional scenario, the pair field changes sign across stripe domain walls, yielding the locking relation 
$\mathbf Q_\text{Stripe}=2\mathbf Q_\text{PDW}$~\cite{AgterbergTsunetsugu2008,Berg2009}.
Large-scale numerical studies of the $t$-$J$ and Hubbard models, based on
density matrix renormalization group (DMRG), quantum Monte Carlo, and related
methods, have established stripe order as one of the most robust
strong-coupling
tendencies~\cite{WhiteScalapino1998,2007PhRvB..76n0505R,2011PhRvL.107r7001L,Corboz2014,Zheng2017,PhysRevB.95.155116,PhysRevLett.127.097003,2022PhRvB.105t5110Z,PhysRevB.107.L220502,JiangKivelson2023,PhysRevLett.132.066002,2024PhRvB.110d5134C,PhysRevB.109.085121,Xu2024,PhysRevB.110.054514},
while finite-$\mathbf Q$ pairing can emerge
concurrently~\cite{2002PhRvL..88k7001H, venderley2019evidence,
  2019PhRvL.122p7001X, peng2021gapless, peng2021precursor, Wietek2022,
  2022npjQM...7...17H, PhysRevB.107.214504, PhysRevB.108.L201110,
  10.3389/femat.2023.1323404, PhysRevB.110.014511,PhysRevLett.133.176501,
  baldelli2025, 2025PNAS..12220963C, 2025CmPhy...8..456Z, gvw1-xk98}.
This raises a central question: \emph{what microscopic magnetic route can
  stabilize finite-momentum pairing in a doped Mott insulator?} Answering this
question requires tracking the competition or cooperation among uniform
pairing, stripe order, and finite-$\mathbf Q$ pairing across tunable
strong-coupling parameters, rather than identifying an oscillatory pair
correlator at a single point.

Altermagnetism, developed through recent symmetry classifications and supported
by materials and spectroscopic studies, provides a symmetry-controlled way to
address this
question~\cite{2004PhRvL..93c6403W,Hayami2019JPSJ,Smejkal2020SciAdv,PhysRevB.102.014422,GonzalezHernandez2021PRL,Smejkal2022PRX,Smejkal2022NRM,Feng2022,Jungwirth2026Nature,2026PhRvB.113a4426Y}.
Like a N\'{e}el antiferromagnet (AFM), an altermagnet consists of two
compensated spin sublattices with zero net magnetization. Unlike a conventional
AFM, however, the two sublattices are related by a crystal rotation rather than
by inversion or translation, allowing spin splitting even without net
magnetization. This distinctive band and symmetry structure has now been
observed or characterized in several altermagnetic (AM)
materials~\cite{Fedchenko2024,Lee2024,Osumi2024MnTe,MnTe2024Nature,Reimers2024CrSbNatCommun,CrSb2024PRL}
and has motivated growing interest in superconductivity in altermagnets and
altermagnet--superconductor
hybrids~\cite{Ouassou2023,Papaj2023,Brekke2023,PhysRevB.110.L060508,2024NatCo..15.1801Z,PhysRevB.110.205120,2025PhRvB.111f4502F,zylh-rqxl,cv8s-tk4c,2026PhRvB.114a4504Z}.
In itinerant systems, finite-momentum Cooper pairing can be driven by
spin-split Fermi
surfaces~\cite{llrq-1k9k,b7rh-v7nq,2026SCPMA..6957011Y,ThermalPDW2026,8cwk-nynj}. In the
Mott limit, however, the AM imprint is encoded instead in bond-anisotropic
exchange and the associated magnon
splitting~\cite{SpinDynamicsRuO22024,kaushal2026}. Correlated AM candidates
such as Ca$_2$RuO$_4$, NiS$_2$, and
CaCrO$_3$~\cite{2023JMMM..58671163C,NiS22025,Ouyang2025}, together with
engineered cold-atom realizations~\cite{PhysRevLett.132.263402}, make this
strong-coupling perspective experimentally motivated in both materials and
synthetic platforms. This distinction makes doped AM Mott insulators a natural
setting for examining how magnetic anisotropy reshapes the competition among
stripe order, uniform pairing, and PDW correlations beyond weak-coupling
Fermi-surface mechanisms.

The checkerboard lattice gives a controlled realization of this symmetry
structure: its two sublattices are related by $C_4$ rotation, and
altermagnetism is encoded microscopically as anisotropic next-nearest-neighbor
exchange~\cite{Tchernyshyov2003,Canals2001,PhysRevLett.132.263402}. This
setting does not introduce generic exchange anisotropy by hand; instead, it
realizes the sublattice-rotation structure of altermagnetism directly in the
microscopic Hamiltonian. This enables a continuous tuning of the AM anisotropy
and provides a controlled way to track how uniform superconductivity, charge
stripes, and finite-$\mathbf Q$ pairing compete or cooperate across doping and
system size.

In this Letter, we investigate whether doping an altermagnetic Mott insulator
can stabilize a PDW. We study the $t$-$J$ model on the checkerboard lattice,
where the two sublattices are related by $C_4$ rotation and altermagnetic
symmetry is encoded microscopically through anisotropic next-nearest-neighbor
hopping and exchange. Using large-scale DMRG, we map the ground-state phase
diagram as a function of doping and altermagnetic anisotropy. With increasing
doping and AM anisotropy, we find that the system evolves from a $d$-wave
superconducting regime to a PDW regime. Most notably, the PDW coexists with
charge stripe order, and their wavevectors obey the unconventional locking
relation $\mathbf Q_{\rm PDW}\approx 2\mathbf Q_{\rm Stripe}$, opposite to the
conventional locking relation.

\begin{figure}[tbp]
  \begin{center}
    \includegraphics[width=0.48\textwidth]{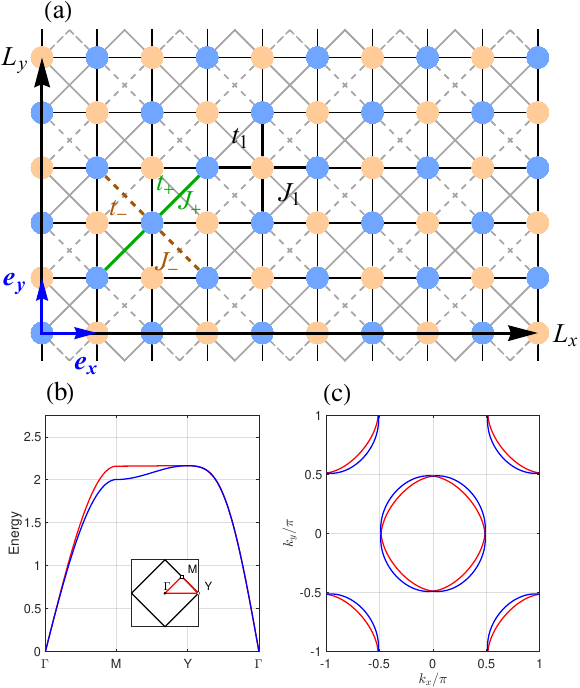}
  \end{center}
  \caption{(a) Checkerboard lattice and model parameters. Blue (orange) circles are $A$ ($B$)
    sublattice sites. Black solid lines: nearest-neighbor bonds with hopping $t_1$ and exchange $J_1$.  Green solid (brown dashed) lines: next-nearest-neighbor bonds with hopping $t_+$ ($t_-$) and anisotropic altermagnetic exchange $J_+$ ($J_-$). (b) Two branches of magnon dispersion along high-symmetry paths and (c) the isoenergy contours at $E=1.8J_1$.  We obtain the magnon dispersion at $\eta=0.7$ by Holstein--Primakoff theory.
  }
  \label{Fig_GeoPhaseDiagram}
\end{figure}

\emph{\color{blue}Model Hamiltonian.}---We study the $t$-$J$ model on the checkerboard lattice with Hamiltonian $H =H_t+H_J$:
\begin{equation}
  \begin{split}
    \!\!  H_t = & -t_1 \sum_{\langle ij\rangle,\sigma}  c^\dagger_{i\sigma}c_{j\sigma}
    - \sum_{\langle\langle ij\rangle\rangle,\sigma}
    t_{ij} c^\dagger_{i\sigma}c_{j\sigma}  + \mathrm{h.c.}                             \\
    \!\!H_J =   & \sum_{\langle ij\rangle}\! J_1 \left(\mathbf{S}_i\cdot\mathbf{S}_j
    - \frac{n_i n_j}{4}\right) \!+\! \sum_{\langle\langle ij\rangle\rangle} \!\! J_{ij}
    \left(\mathbf{S}_i\cdot\mathbf{S}_j - \frac{n_i n_j}{4}\right),
  \end{split}
  \label{eq:tJ}
\end{equation}
where $c_{i\sigma}$ annihilates a fermion at site $i=(x,y)$ with spin $\sigma$, $\mathbf {S}_i$ is the spin operator, and $n_i$ denotes the electron-number operator. The Hilbert space is constrained by the
no-double-occupancy condition. The checkerboard lattice in
Fig.~\ref{Fig_GeoPhaseDiagram}(a) has nearest-neighbor bonds with hopping
$t_1$ and exchange $J_1$. On next-nearest-neighbor (NNN) bonds, the hopping and
exchange take bond-dependent values $ t_{\pm} = t_2(1\pm \eta)$ and $ J_{\pm} =
  -({t_{\pm}}/{t_1})^2 J_1$.
This bond-anisotropic exchange realizes the altermagnetic sublattice relation:
the two compensated spin sublattices are connected by a $90^\circ$ crystal
rotation rather than by inversion or translation. We focus on the parameter regime $0.6
  \leq \eta \leq 0.8$ to control the AM anisotropy, and fix $t_1=3J_1=3$, $t_2/t_1
  = 0.5$ throughout while varying hole doping $\delta$ and AM anisotropy $\eta$.

\begin{figure*}[t]
  \begin{center}
    \includegraphics[width=0.9\textwidth]{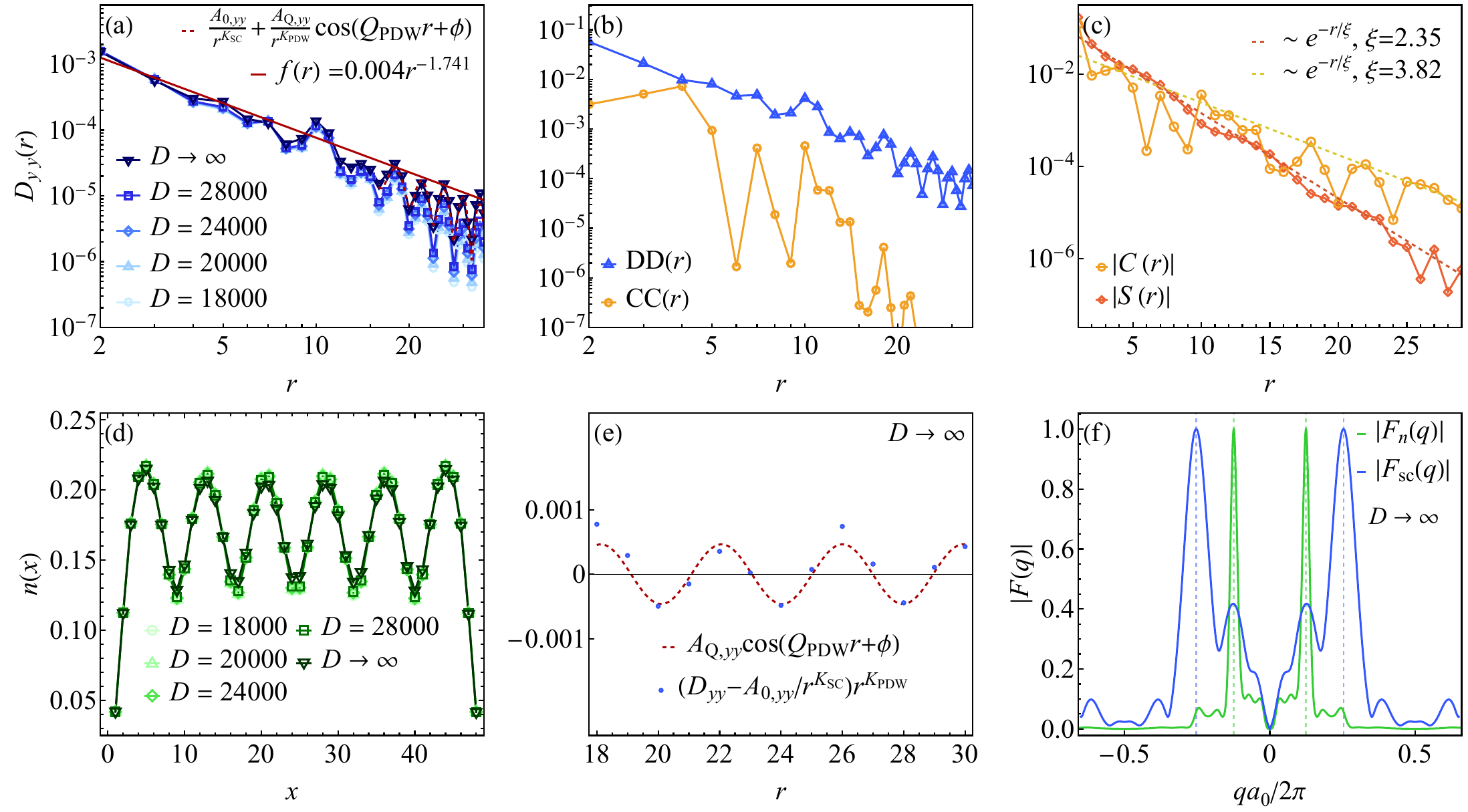}
  \end{center}
  \caption{
  DMRG results for the PDW phase on an $N=48\times 6$ cylinder at doping $\delta=1/6$ and AM anisotropy $\eta=0.7$.
  (a) Pair correlation $D_{yy}(r)$ as a function of distance $r$ for several bond dimensions $D$, together with the extrapolation to $D\to\infty$.
  The red solid line shows a power-law fit $f(r)=0.004r^{-1.741}$.
  The long-distance behavior (red dashed line) follows the two-scale form of Eq.~\eqref{eq:D_scpdw} with
  $K_\mathrm{SC}=2.30 $ and $K_\mathrm{PDW}=1.40$.
  (b) Normalized superconducting correlations $\mathrm{DD}(r)\equiv D(r)/\delta^2$ (blue triangles) and single-particle Green's function $\mathrm{CC}(r) \equiv [C(r) / \delta]^2$ (orange circles) on a log--log scale.
  (c) The single-particle Green's function $C(r)$
  and the spin correlation $S(r)$ on a semi-log scale. (d) Local hole density
  $n(x)$, showing stripe modulation with a period $\approx 8a_0$. (e) Extracted
  PDW component, displaying sign-changing oscillations. The regular sign changes,
  together with the equal weights of the $\pm Q_{\mathrm{PDW}}$ peaks in
  $|F_{\mathrm{sc}}(q)|$, consistent with the Larkin--Ovchinnikov (LO) standing-wave modulation. (f) Structure factors: $|F_{\mathrm{sc}}(q)|$ (blue) of
  the pairing correlators and $|F_{n}(q)|$ (green) of charge density for bond
  dimension $D\to\infty$. Vertical dashed lines indicate the characteristic
  momenta $Q_{\mathrm{PDW}}\approx 0.25(2\pi/a_0)$ and
  $Q_{\mathrm{Stripe}}\approx 0.12(2\pi/a_0)$. The peak heights are scaled to 1
  for better visibility and comparison. } \label{Fig_dt07N48x6}
\end{figure*}

We employ DMRG on cylinders of size $L_x\times L_y$ with periodic (open)
boundary conditions along the $\hat{y}$ ($\hat{x}$) direction [see
    Fig.~\ref{Fig_GeoPhaseDiagram}(a)], studying widths $L_y=6$ and $8$ with $L_x$
up to 48. We keep bond dimensions up to $D=36{,}000$ with $SU(2)$ spin
symmetry, equivalent to $D\approx100{,}000$ states in the $U(1)$ basis. The
ground-state order is characterized through the local hole density $ n(x)
  \equiv 1-\sum_{y,\sigma} \langle
  c^\dagger_{(x,y)\sigma}c^{\phantom{\dagger}}_{(x,y)\sigma}\rangle/L_y, $ and
the pair correlator
\begin{equation}
  D_{\alpha\beta}(r) =
  \langle\Delta^\dagger_\alpha(x_0,y_0)\,
  \Delta^{\phantom{\dagger}}_\beta(x_0+r,y_0)\rangle,
\end{equation}
where $\Delta_\alpha(x,y)\equiv\frac{1}{\sqrt{2}}\sum_\sigma\sigma\,
  c_{(x,y)\sigma}\,c_{(x,y)+\mathbf{e}_\alpha,\bar \sigma}$ is the singlet pair
operator on the bond $\alpha$.

\emph{\color{blue}Altermagnetic spin wave.}---At half filling, the spin-wave spectrum of the AM checkerboard lattice encodes the full symmetry structure of the model and provides a direct spectroscopic fingerprint of the altermagnetic order. Switching on $\eta > 0$ breaks the degeneracy of the magnon branches \cite{PhysRevLett.134.196701}, as depicted in Fig.~\ref{Fig_GeoPhaseDiagram}(b), based on Holstein--Primakoff theory.
The split magnon branches carry opposite spin character in the four
Brillouin-zone quadrants, manifested by the splitting along the high-symmetry
line $\Gamma \rightarrow M \rightarrow Y $. The two magnon branches are related
by the $\mathcal{T}\cdot C_4$ symmetry and exhibit a $d_{xy}$-wave spin
splitting $\propto \sin k_x \sin k_y$. The preserved $\mathcal{T}\cdot C_4$
symmetry therefore imposes a degeneracy on nodal lines at the spin-wave level
in Fig.~\ref{Fig_GeoPhaseDiagram}(c), even though $\mathcal{T}$ and $C_4$ are
individually broken. This hallmark distinguishes the AM magnon spectrum from
that of a conventional N\'eel antiferromagnet. The ferromagnetic NNN exchanges
introduce an anisotropy of the AFM background that intensifies with $\eta$.
This motivates a systematic examination of how the phase diagram evolves with
doping and AM anisotropy.

\emph{\color{blue}DMRG results on PDW.}---
We first characterize the unconventional PDW state at large altermagnetic anisotropy.
Figure~\ref{Fig_dt07N48x6} shows results for an $N=48\times6$ cylinder at doping $\delta=1/6$ and AM anisotropy  $\eta=0.7$.
Overall, the pair correlator $D_{yy}(r)$ follows a power-law decay with fitting function $f(r)=0.004 r^{-1.741}$ after $D\to\infty$ extrapolation, shown in Fig.~\ref{Fig_dt07N48x6}(a).
Moreover, the renormalized pair correlator $\mathrm{DD}(r)\equiv D(r)/\delta^2$ decays algebraically and dominates over the renormalized single-particle propagator
$\mathrm{CC}(r) \equiv [C(r) / \delta]^2$
from the single-particle Green's function $C(r)= \sum_\sigma \langle     c_{(x_0,y_0)\sigma}^\dagger c_{(x_0+r,y_0)\sigma}^{}\rangle$ [see Fig.~\ref{Fig_dt07N48x6}(b)].
The spin correlation $S(r)= \langle\mathbf{S}_{(x_0,y_0)}\cdot \mathbf{S}_{(x_0+r,y_0)}^{}\rangle$ [see Fig.~\ref{Fig_dt07N48x6}(c)]  decays exponentially, indicating a short-range magnetic background consistent with the properties of the PDW phase in the $t$-$J$ model~\cite{Berg2009,AgterbergTsunetsugu2008}.

To resolve the pairing structure from the finite-size calculations, we fit the
asymptotic form at large distance
\begin{equation}
  D_{\alpha\beta}(r)
  \rightarrow \frac{A_{0,\alpha\beta}}{r^{K_\mathrm{SC}}}+
  \frac{A_{Q,\alpha\beta}\cos(Q_\mathrm{PDW} r + \phi)}{r^{K_\mathrm{PDW}}},
  \label{eq:D_scpdw}
\end{equation}
where the first term captures residual uniform superconducting (SC)
contributions, and the second term represents the PDW component with wavevector
$Q_\mathrm{PDW}$. At long distances, the PDW component decays more slowly
according to the extracted exponents $K_\mathrm{SC}>2>K_\mathrm{PDW}$ from the
numerical fitting \cite{note}. At short distances, the fitted uniform component
has a substantially larger amplitude, with $|A_0|/|A_Q|\approx 29.0$.
Nevertheless, because $K_{\mathrm{PDW}}<K_{\mathrm{SC}}$, the relative weight
of the PDW component grows algebraically with distance and is expected to
dominate asymptotically. 
The ratios $A_{0,xy}/A_{0,xx}<0$ and
$A_{0,xy}/A_{0,yy}< 0$ indicate an underlying $d$-wave character of local
pairing.
This reveals a two-scale pairing structure: a short-distance uniform $d$-wave
component coexists with a more slowly decaying finite-momentum PDW component.

Simultaneously, charge stripe order coexists with the PDW correlations. In
Fig.~\ref{Fig_dt07N48x6}(d), the hole density $n(x)$ exhibits a robust stripe
modulation with period $\sim 8a_0$ and amplitude $\delta n\approx 0.1$,
corresponding to $Q_\text{Stripe}\approx 0.12(2\pi/a_0)$. Further, detailed
features are depicted in Fig.~\ref{Fig_dt07N48x6}(f) by looking at the
structure factors $F(q)$ of the pair correlators and charge density, which
reveal an unconventional wave-vector locking. 
To characterize the finite-momentum pairing component, we define $ F_\text{sc}(q) = \sum_{r=r_\text{min}}^{r_\text{max}}   (g(r)-\bar{g}) e^{iqr}$, where $g(r)=(D_{yy}(r)-A_{0,yy}r^{-K_\mathrm{SC}})/(A_{Q,yy}r^{-K_\mathrm{PDW}})$.
Similarly, we obtain the structure factor $F_n(q)$ of the charge density
distribution. In Fig.~\ref{Fig_dt07N48x6}(f), the dominant peaks of
$|F_\text{sc}(q)|$ occur at $Q_\text{PDW}\approx \pm 0.25(2\pi/a_0)$. The
secondary peaks coincide with the peaks of $|F_{n}(q)|$ at
$Q_\text{Stripe}\approx \pm 0.12(2\pi/a_0)$, giving $Q_\text{PDW}\approx
  2Q_\text{Stripe}$.

\begin{figure}[t]
  \begin{center}
    \includegraphics[width=0.49\textwidth]{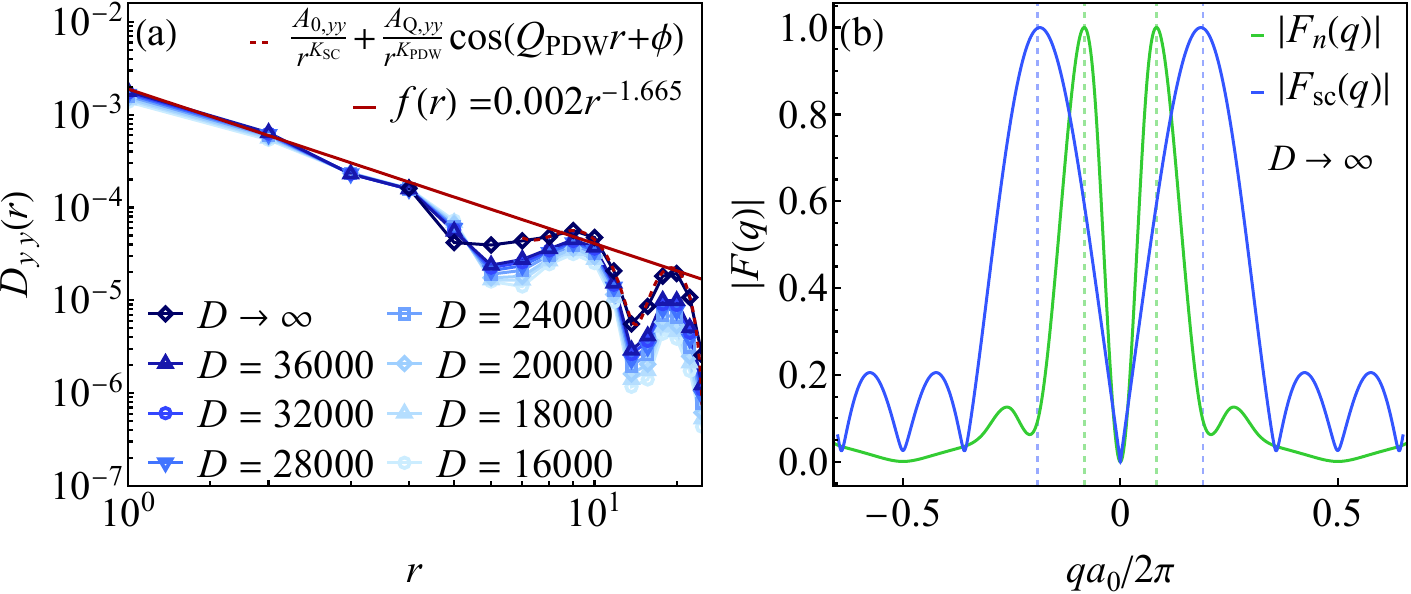}
  \end{center}
  \caption{DMRG results for the PDW  phase on an $N=24\times 8$ cylinder at $\delta=1/8$ and $\eta=0.8$. (a) Pair correlation $D_{yy}$ extrapolated to $D\to\infty$. The red solid line denotes a single power-law fit, $f(r)=0.002r^{-1.665}$, while the red dashed line denotes the two-scale fit of Eq.~\eqref{eq:D_scpdw}, with $K_{\rm SC}=2.2$  and $K_{\rm PDW}=1.4$~\cite{note2}.
  (b) Structure factors: superconducting structure factor $|F_{\mathrm{sc}}(q)|$ (blue) and density structure factor $|F_{n}(q)|$ (green) after extrapolation to $D\to\infty$.  $|F_{\mathrm{sc}}(q)|$ exhibits a peak at $Q_{\mathrm{PDW}}\approx 0.19(2\pi/a_0)$, while the density structure factor $|F_{n}(q)|$ peaks at $Q_{\mathrm{Stripe}}\approx 0.10(2\pi/a_0)$. The peak heights are scaled to 1 for better visibility and comparison.
  }
  \label{Fig_Delta18dt08N24x8}
\end{figure}

To test the robustness of these observations, we also study a wider cylinder
$N=24\times8$ at $\delta=1/8$ and $\eta=0.8$, as shown in
Fig.~\ref{Fig_Delta18dt08N24x8}(a), deep in the PDW regime. The overall
extrapolated pair correlation follows $D_{yy}(r) \approx 0.002r^{-1.665}$, well
converged up to $D=36{,}000$, with slightly slower decay than in the
$48\times6$ system. The two-scale form of Eq.~\eqref{eq:D_scpdw} shows
$A_{0,yy}=12.8A_{Q,yy}=8.9\times10^{-3}, K_{\rm SC}=2.2$, and $K_{\rm
      PDW}=1.4$, indicating a PDW component that decays more slowly than the uniform
component and is therefore expected to dominate asymptotically. The
superconducting structure factor peaks at $Q_\text{PDW}\approx 0.19(2\pi/a_0)$,
while the density modulation peaks at $Q_\text{Stripe}\approx
  0.10(2\pi/a_0)\approx Q_\text{PDW}/2$ [see Fig.~\ref{Fig_Delta18dt08N24x8}(b)].

\begin{figure}[t]
  \begin{center}
    \includegraphics[width=0.32\textwidth]{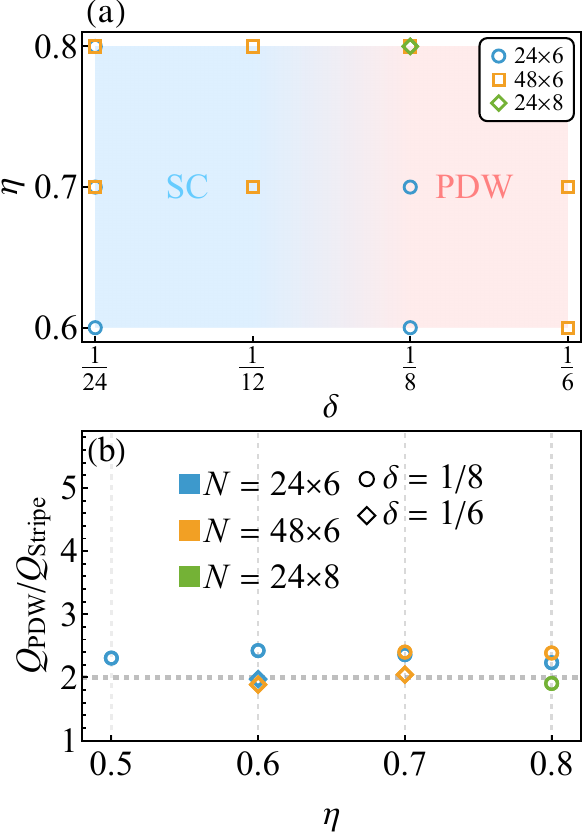}
  \end{center}
  \caption{
    (a) Ground-state phase diagram as a function of hole doping $\delta$ and AM  anisotropy $\eta$. We identify two regimes: coexistence of SC and CDW (blue) and coexistence of PDW and stripe (pink). (b) Wave-vector locking in the PDW phase: $\mathbf Q_\mathrm{PDW}\approx 2\mathbf Q_\mathrm{Stripe}$ (dashed line) across various system sizes and parameters. Colors denote system sizes $N$, while symbols denote different hole doping concentrations $\delta$.
  }
  \label{Fig4_PhaseDiagram}
\end{figure}

\emph{\color{blue}Phase diagram.}---Figure~\ref{Fig4_PhaseDiagram}(a) shows the ground-state phase diagram versus hole doping $\delta$ and AM anisotropy $\eta$. We identify two regimes by comparing superconducting and density structure factors. In the SC regime (blue), the pair correlation shows power-law decay with a  charge-density-wave (CDW) modulation at wavevector $Q_\text{CDW}$.
In the PDW regime (pink), the pair correlation exhibits a finite-momentum
oscillation with wavevector $Q_\text{PDW}$, while $F_\text{sc}(q)$ is dominated
by a finite-momentum peak at $Q_\mathrm{PDW}>0$ and the local hole density and
$|F_{n}(q)|$ show stripe modulation at $Q_\text{Stripe}\approx Q_\text{PDW}/2$.
Within the explored parameter window, this regime appears for $\delta \gtrsim
  1/8$ and $0.8 \gtrsim \eta \gtrsim 0.6$. Such unconventional wave-vector
locking has not, to our knowledge, been reported in the conventional $t$-$J$
model or in related deformations on other lattice geometries lacking AM
anisotropy. This suggests that AM might be one of the key ingredients in
promoting the intertwined PDW and stripe states within the explored parameter
regime.

\emph{\color{blue}Discussion.}---
The main result of our study is that, within the explored parameter window,
the balance between uniform $d$-wave SC and PDW correlations changes as
$\eta$ and doping are varied. The observed wave-vector locking $\mathbf Q_{\text{PDW}} \approx
  2\mathbf Q_{\text{Stripe}}$ along the cylinder direction is the reverse of the stripe scenario proposed
for cuprates~\cite{Berg2009,RevModPhys.87.457},
and is also distinct from the weak-coupling Fulde--Ferrell--Larkin--Ovchinnikov (FFLO)
picture~\cite{Fulde1964,LarkinOvchinnikov1964}.  In
Fig.~\ref{Fig4_PhaseDiagram}(b), we compile the DMRG data obtained for the parameter sets studied in this work, illustrating the consistency of this locking across system
sizes and dopings.

In the PDW phase, the uniform $d$-wave component decays more rapidly at long
distances than the PDW component. The pair correlators nevertheless reveal a
sizable short-range uniform component, indicating that strong local Cooper
pairing survives, while the corresponding uniform pairing susceptibility
remains finite because $K_{\mathrm{SC}}>2$.
The observed wave-vector locking can be rationalized at the symmetry level
within a Ginzburg--Landau framework involving the stripe field $\rho_{\mathbf
    Q_\mathrm{Stripe}}$, the uniform superconducting field $\Delta_0$, and the PDW
fields $\Delta_{\pm\mathbf Q_\mathrm{PDW}}$. Two symmetry-allowed couplings are
particularly relevant. The first is a linear stripe--PDW coupling
\cite{Agterberg2020}, $ \rho_{\mathbf Q_\mathrm{Stripe}} \Delta^*_{\mathbf
    Q_\mathrm{PDW}} \Delta_{-\mathbf Q_\mathrm{PDW}} +\mathrm{c.c.}, $ which favors
the conventional locking $\mathbf Q_\mathrm{PDW}=\mathbf Q_\mathrm{Stripe}/2$.
The second is a quadratic stripe coupling, $ (\rho_{\mathbf
    Q_\mathrm{Stripe}}^*)^2 \Delta_0^* \Delta_{\mathbf Q_\mathrm{PDW}}
  +\mathrm{c.c.}, $ which instead couples the uniform superconducting component
to a PDW at $\mathbf Q_\mathrm{PDW}=2\mathbf Q_\mathrm{Stripe}$, which is
consistent with the observed locking.
Once stripe order is established ($\rho_{\mathbf Q_\mathrm{Stripe}}\neq 0$),
this term directly mixes $\Delta_0$ and $\Delta_{\mathbf Q_\mathrm{PDW}}$.
Together with the fitted power-law decay exponents,
$K_\mathrm{PDW}<2<K_\mathrm{SC}$, this suggests that the PDW channel at
$2\mathbf Q_\mathrm{Stripe}$ is favored over the conventional $\mathbf
  Q_\mathrm{Stripe}/2$ channel. In this regime, the dominant PDW correlations
coexist with a short-range uniform pairing component.
This behavior may also be related to the nematic spin-nematic state discussed in Ref.~\cite{2014PhRvB..89p5126S}, where a $d$-wave PDW is predicted, consistent with the local pairing symmetry inferred from our DMRG results, within the limitations imposed by the accessible cylinder geometry.
The precise microscopic
mechanism by which the AM background selects the $2\mathbf Q_\mathrm{Stripe}$
channel over the conventional $\mathbf Q_\mathrm{Stripe}/2$ channel, likely
tied to how the AM-anisotropic NNN exchange modifies the pairing
susceptibility, remains an open question, calling for future theoretical work.

\emph{\color{blue}Conclusion.}---
We have studied the doped checkerboard $t$-$J$ model as a minimal
strong-coupling realization of an AM Mott insulator. By varying hole doping
$\delta$ and AM anisotropy $\eta$, large-scale DMRG reveals two regimes in the
explored parameter window: a uniform $d$-wave superconducting regime  and a PDW regime.
The central result is the reversed wave-vector locking $\mathbf
  Q_\text{PDW}\approx 2\mathbf Q_\mathrm{Stripe}$ along the cylinder direction in
the PDW phase where charge stripe order coexists, opposite to the conventional
relation $\mathbf Q_\text{Stripe}=2\mathbf Q_\mathrm{PDW}$.

The results therefore position AM Mott systems as a useful route to
unconventional PDW physics in the strong-coupling regime. Candidate directions
include correlated AM materials such as Ca$_2$RuO$_4$, NiS$_2$, CaCrO$_3$,
YVO$_3$,  LaTiO$_3$, $\mathrm{CsV}_2 \mathrm{Se}_2 \mathrm{O}$~\cite{2023JMMM..58671163C,2024PhRvM...8f4403M,Cuono2024NiS2,NiS22025,Ouyang2025,FuDaran2026,LiYuanzhe2026},
where pressure, strain, doping, or proximity engineering may tune the magnetic
and electronic structure, as well as ultracold atom
realizations~\cite{PhysRevLett.132.263402} where the exchange anisotropy and
doping can be controlled directly. These platforms offer complementary ways to
test whether altermagnetic exchange anisotropy can promote finite-momentum
superconductivity beyond the specific model studied here.

\begin{acknowledgments}
  We thank Roderich Moessner and Alexander Wietek for helpful discussions. This work was supported by the National Natural Science Foundation of China (Grant Nos. 92477106 and 12504184) and the Fundamental Research Funds for the Central Universities.
\end{acknowledgments}


\begin{thebibliography}{86}
\bibitem [{not({\natexlab{a}})}]{note}%
  \BibitemOpen
  \bibinfo {note} {We fit $D_{yy}(r)=A_{0,yy}r^{-K_{\rm SC}}
  +A_{Q,yy}\cos(Q_{\rm PDW}r+\phi)r^{-K_{\rm PDW}}$ over $10\le r\le 31$. The
  wavevector is fixed to the peak in the Fourier spectrum of the
  envelope-flattened correlator, $Q_{\rm PDW}\simeq 0.25\times(2\pi/a_0)$,
  corresponding to a period of approximately $4a_0$; allowing $Q_{\rm PDW}$ to
  vary changes its fitted value by less than $2\%$. All quoted intervals are
  $95\%$ confidence intervals. We obtain $K_{\rm SC}=2.30\pm0.25$ and $K_{\rm
  PDW}=1.4^{+0.7}_{-0.8}$. The inequality $K_{\rm PDW}<K_{\rm SC}$ is supported
  at the $99.9\%$ confidence level, while the stronger condition $K_{\rm
  PDW}<2<K_{\rm SC}$ holds at approximately $96\%$ confidence.}\BibitemShut
  {Stop}%
\bibitem [{not({\natexlab{b}})}]{note2}%
  \BibitemOpen
  \bibinfo {note} {We fit $D_{yy}(r)=A_{0,yy}r^{-K_{\rm SC}}
  +A_{Q,yy}\cos(Q_{\rm PDW}r+\phi)r^{-K_{\rm PDW}}$ over $7\le r\le 15$. The
  extracted wavevector, $Q_{\rm PDW}\simeq 0.19\times(2\pi/a_0)$, corresponding
  to a period of approximately $5.3a_0$, is independently confirmed by the
  Fourier spectrum of the envelope-flattened correlator. All quoted intervals
  are $95\%$ confidence intervals. We obtain $K_{\rm PDW}=1.4^{+0.3}_{-0.4}$
  and $K_{\rm SC}=2.2^{+0.3}_{-0.1}$. The inequality $K_{\rm PDW}<K_{\rm SC}$
  is supported at nearly the $100\%$ confidence level.}\BibitemShut {Stop}%
\end{thebibliography}
\end{document}